\documentclass[epj]{webofc}
\usepackage[utf8]{inputenc}
\usepackage[varg]{txfonts}
\usepackage{booktabs}
\usepackage{xcolor}
\definecolor{darkred}{rgb}{0.4,0.0,0.0}
\definecolor{darkgreen}{rgb}{0.0,0.4,0.0}
\definecolor{darkblue}{rgb}{0.0,0.0,0.4}
\usepackage[bookmarks,linktocpage,colorlinks,
    linkcolor = darkred,
    urlcolor  = darkblue,
    citecolor = darkgreen]{hyperref}
\wocname{EPJ Web of Conferences}
\woctitle{Lattice2017}

\usepackage{slashed}
\newcommand{\cA}{\ensuremath{\mathcal A} }

\newcommand{\Cbb}{\ensuremath{\mathbb C} }
\newcommand{\cgrav}{\ensuremath{c_{\text{grav}}} }
\newcommand{\cD}{\ensuremath{\mathcal D} }
\newcommand{\cDb}{\ensuremath{\overline{\mathcal D}} }
\newcommand{\cF}{\ensuremath{\mathcal F} }
\newcommand{\cFb}{\ensuremath{\overline{\mathcal F}} }
\newcommand{\Ibb}{\ensuremath{\mathbb I} }
\newcommand{\cN}{\ensuremath{\mathcal N} }
\newcommand{\vn}{\mathbf{n}}
\newcommand{\cO}{\ensuremath{\mathcal O} }
\newcommand{\cP}{\ensuremath{\mathcal P} }
\newcommand{\cQ}{\ensuremath{\mathcal Q} }
\newcommand{\cU}{\ensuremath{\mathcal U} }
\newcommand{\cUb}{\ensuremath{\overline{\mathcal U}} }
\newcommand{\al}{\ensuremath{\alpha} }
\newcommand{\ga}{\ensuremath{\gamma} }

\newcommand{\lam}{\ensuremath{\lambda} }
\newcommand{\lalat}{\ensuremath{\lam_{\text{lat}}} }
\newcommand{\glN}{\ensuremath{\mathfrak{gl}(N, \Cbb)} }
\newcommand{\Tr}[1]{\ensuremath{\text{Tr}\left[ #1 \right]} }
\newcommand{\vev}[1]{\ensuremath{\left\langle #1 \right\rangle} }

\begin{document}
\selectlanguage{english}
\title{Testing the holographic principle using lattice simulations}
\author{
  \firstname{Raghav G.} \lastname{Jha}\inst{1}\thanks{Speaker, \email{rgjha@syr.edu}}   \and
  \firstname{Simon} \lastname{Catterall}\inst{1} \and
  \firstname{David}  \lastname{Schaich}\inst{2} \and
  \firstname{Toby}  \lastname{Wiseman}\inst{3}
}

\institute{
  Department of Physics, Syracuse University, Syracuse, New York 13244, United States \and
  AEC Institute for Theoretical Physics, University of Bern, 3012 Bern, Switzerland \and
  Theoretical Physics Group, Blackett Laboratory, Imperial College, London SW7 2AZ, United Kingdom
}

\abstract{
  The lattice studies of maximally supersymmetric Yang-Mills (MSYM) theory at strong coupling and large $N$ is important for verifying gauge/gravity duality.
  Due to the progress made in the last decade, based on ideas from topological twisting and orbifolding, it is now possible to study these theories on the lattice while preserving an exact supersymmetry on the lattice.
  We present some results from the lattice studies of two-dimensional MSYM which is related to Type II supergravity.
  Our results agree with the thermodynamics of different black hole phases on the gravity side and the phase transition (Gregory--Laflamme) between them.
}

\maketitle

\section{Introduction}\label{intro}
MSYM theories in $p + 1$ dimensions has been conjectured to provide a holographic description of string theories containing D$p$-branes.
The duality states that ($p + 1$)-dimensional SYM with gauge group SU($N$) is dual to a Type~IIA (even $p$) or Type~IIB (odd $p$) superstring containing $N$ coincident D$p$-branes in the `decoupling' limit~\cite{Itzhaki:1998dd, Aharony:1999ti, Maldacena:1997re}.
In this proceedings, we present results from the lattice simulation of maximally supersymmetric SYM in two dimensions at finite temperatures, with the spatial circle direction compactified, i.e. on a Euclidean 2-torus.
Generally, the 2-torus is taken to be rectangular but we emphasize here that it can also be skewed allowing us to test the duality for a wider class of theories.
For our detailed recent work on the two-dimensional theory, on which this proceedings is based, see Ref.~\cite{Catterall:2017lub}.

At large~$N$ and low temperatures, the dual string theory is well described by supergravities whose dynamics are given by certain charged black holes.
Two classes of black holes are required to describe these dynamics---those that wrap the spatial circle (so called `homogeneous black strings') and those that are localized on it (`localized black holes').
The two classes have different thermodynamic behaviours, and there is a first-order dual deconfinement transition associated with the spatial cycle of the torus (Wilson loop being the order parameter).
It should be possible to reproduce these from the thermal physics of the SYM in the strong coupling regime if the gauge/gravity duality is correct.
It is conjectured that this transition extends to high temperatures where the phase structure has been revealed from numerical and analytic treatments.
The emergence of gauge/gravity duality has meant that whilst SYM provides a fundamental and microscopic quantum description of certain gravity systems, there still is no complete derivation of the dual black hole behavior from ($p + 1$)-dimensional SYM.

It is natural to make use of lattice field theory, which is well suited to study the thermodynamics of strongly coupled systems to understand this intricate relation in detail.
Several works over the past decade have studied the thermal behaviour of the $p = 0$ SYM quantum mechanics and striking agreement has been seen~\cite{Anagnostopoulos:2007fw, Catterall:2008yz, Hanada:2008gy, Hanada:2008ez, Hanada:2016zxj, Berkowitz:2016jlq}.
However, the dual gravity in that setting is simpler than in the $p = 1$ case we focus on here, where there are different black holes to probe, a gravity phase transition to observe, and the shape of the torus to vary.
Previous lattice investigations~\cite{Catterall:2010fx} probed $p = 1$ MSYM using small $N = 3$ and 4, finding evidence for consistency with gravity predictions.

In the last few years, we have made progress in the lattice studies of the $p = 3$ theory, $\cN = 4$ SYM, using a novel construction based on discretization of a topologically twisted form of the continuum $\cN = 4$ action.
See Ref.~\cite{Catterall:2009it} for a review of this approach.
The upshot of this new lattice construction is that it preserves a closed subalgebra of the supersymmetries at non-zero lattice spacing.
Numerical studies of the four-dimensional theory are in progress~\cite{Catterall:2014vka, Schaich:2014pda, Catterall:2014vga, Catterall:2015ira, Schaich:2015daa, Schaich:2015ppr, Schaich:2016jus}, but are very expensive because of the large number of degrees of freedom.
In this regard, lower-dimensional theories are more tractable and can be well-studied at large $N$.

The recent lattice constructions are based on non-hypercubic Euclidean lattices, which when made periodic are naturally adapted to skewed tori.
We dimensionally reduce an $\cN = 4$ lattice system to give a discretization of two-dimensional SU($N$) SYM on an $A_2^*$ lattice, preserving four exact supercharges at non-zero lattice spacing.
After applying appropriate BCs we then carry out calculations for $N \le 16$, large enough to see dual gravity behaviour.
We see both phases of dual black hole behaviour in the appropriate low temperature regime, and nice agreement between the generalized SYM thermodynamics and that predicted by gravity.
We also see a transition between these phases, again compatible with the expectation from gravity, which extends to high temperature as expected.

\section{\label{sec:gravity}Thermal large $N$ $(1 + 1)$-dimensional SYM on a circle}
We review the predictions for large-$N$ $p = 1$ SYM, compactified on a circle of size $L$ at temperature $T = 1 / \beta$.
From a Euclidean perspective this theory lives on a flat rectangular 2-torus, with side lengths $\beta$ and $L$.
Although we consider the theory on a \emph{skewed} torus with a non-zero skewing parameter $\ga \ne 0$, here it will suffice to consider the rectangular torus case.
More details can be found in Ref.~\cite{Catterall:2017lub}.
The Euclidean action of the theory is
\begin{equation}
  \label{eq:SYMaction}
  S = \frac{N}{4\lam} \int d\tau\, dx\ \Tr{F_{\mu\nu}F^{\mu\nu} + 2\left(D_{\mu} X^I\right)^2 - \left[X^I, X^J\right]^2 + \bigg[\Psi \left(\slashed{D} - \left[\Gamma^I X^I, \, \cdot \,\right]\right) \Psi \bigg]}.
\end{equation}
Here $X^I$ with $I = 2, \ldots, 9$ are the eight spacetime scalars representing the transverse degrees of freedom of the branes.
They are $N\times N$ hermitian matrices in the adjoint representation of the gauge group.
The fermion $\Psi$ and matrices $\Gamma^I$ descend from a dimensional reduction of a ten-dimensional Euclidean Majorana--Weyl spinor, with $\Psi$ also transforming in the adjoint.
The dimensionful 't~Hooft coupling $\lam = N g_{YM}^2$ may be used to construct two dimensionless quantities that control the dynamics: $r_{\beta} = \beta \sqrt{\lam}$ and $r_L = L \sqrt{\lam}$.
We define the dimensionless temperature $t = 1 / r_{\beta}$.
Since we are interested in the large-$N$ 't~Hooft limit we wish to consider $N \to \infty$ with $r_{\beta}$ and $r_L$ fixed.
The main observables we consider are thermodynamic quantities related to the expectation value of the bosonic action, and also the Wilson loop $P_{\beta}$ and $P_L$, where
\begin{equation}
  \label{eq:Pdefn}
  P_{\beta, L} = \frac{1}{N} \vev{\left| \Tr{\cP e^{i\oint_{\beta, L} A}} \right|}
\end{equation}
which wrap about the Euclidean thermal circle and spatial circle of the two-dimensional Euclidean torus respectively.
For the large-$N$ theory these act as order parameters for phase transitions associated to breaking of the $Z_N$ center symmetry of the gauge group.
Since we are in a finite volume, we can only have a phase transition at large $N$.
For the thermal circle this is the usual thermal deconfinement transition, with vanishing Polyakov loop $P_{\beta} = 0$ at large~$N$ indicating the (unbroken) confined phase, and $P_{\beta} \ne 0$ being the (broken) deconfined phase.
We will use similar terminology for $P_L$, namely that $P_L \ne 0$ indicates `deconfined' spatial behaviour while $P_L = 0$ corresponds to `confined' spatial behaviour.

\section{\label{sec:dualGrav}Type IIA/IIB supergravity}
At large~$N$ and low temperatures $r_{\beta} \gg 1$, holography predicts a gravity dual given by D1-charged black holes in IIB supergravity~\cite{Itzhaki:1998dd}.
Large~$N$ is required to suppress string quantum corrections to the supergravity.
In order to suppress $\al'$ corrections we require $1 \ll r_{\beta}$, and to avoid winding mode corrections about the circle we need $r_{\beta} \ll r_L^2$.
When $r_{\beta} \sim r_L^2$ the IIB solution is unstable due to string winding modes on the spatial circle.
We can use T-duality to obtain a IIA gravity solution, which reproduces the IIB solution in the regime $r_L \ll r_{\beta} \ll r_L^2$ where the latter exists and describes the physics.
The IIA solution also covers smaller circle sizes all the way down to the limit $r_L \to 0$ where the physics is that of the dimensionally reduced supersymmetric quantum mechanics.

The above solution is homogeneous on the circle---a `homogeneous black string'.
Being related by T-duality, under which the action is invariant, it has precisely the same thermodynamics as the IIB solution above.
Namely it predicts the thermodynamic behaviour,
\begin{equation}
  \label{eq:D1phase}
  \frac{s_{\text{Bos}}}{N^2 \lam} = -\frac{2^3 \pi^{\frac{5}{2}}}{3^4} t^3 \simeq -1.7275 t^3
\end{equation}
for the SYM bosonic action density $s_{\text{Bos}}$, with $t = 1 /r_{\beta}$ the dimensionless temperature.\footnote{The skewing parameter is $\ga = -1 / 2$ for the $A_2^{\star}$ lattice we use.  Since we do not have a rectangular geometry ($\ga = 0$), the thermodynamic potentials are not well defined and we avoid discussing the free energy.}
However, the winding mode in the original IIB frame is now a classical Gregory--Laflamme (GL) instability in this IIA frame.
One finds the above solution is dynamically unstable when $r_L^2 \le c_{\text{GL}} r_{\tau}$, with $c_{\text{GL}} \simeq 2.24$.
Thus at smaller circle sizes the above solution remains, but it is not the relevant one for the dynamics, which instead is given in terms of a `localized black hole' solution.

The localized black hole (small black holes that break the asymptotic $S^1$ translation invariance at some particular position on the $\beta$-cycle) solutions are not known analytically.
However, recently a numerical construction of these solutions has been performed~\cite{Dias:2017uyv}.
At $r_L^2 = \cgrav r_{\beta} \simeq 2.45 r_{\beta}$, Ref.~\cite{Dias:2017uyv} found evidence for a first-order phase transition to the localized phase, which then dominates the homogeneous one for smaller $r_L^2 / r_{\beta}$, having lower free energy density.

Even though the analytic form of these localized solutions is not known generally, they simplify in the limit when the horizon is small compared to the circle size.
In SYM variables, when $r_L^2 t \ll 1$, the solutions have an approximate localized behaviour,
\begin{equation}
  \label{eq:D0phase}
  \frac{s_{\text{Bos}}}{N^2 \lam} = -2.469 \frac{t^{16 / 5}}{\al^{2 / 5} (1 - \ga^2)^{7 / 5}}.
\end{equation}
The approximate behaviour of the localized phase \eqref{eq:D0phase}, is only valid for $r_L^2 t \ll 1$.
However, the numerical solutions~\cite{Dias:2017uyv} show that this approximation remains reasonable throughout the range where the localized phase dominates the canonical ensemble.
We will refer to the homogeneous phase as the \emph{D1~phase}, since in the IIB duality frame it is the D1-brane solution.
It may also be seen as a homogeneous D0-brane solution in the IIA frame.
We will refer to the localized phase as the \emph{D0~phase}, since it may \emph{only} be seen in gravity in the IIA frame where it is a localized D0-brane black hole.

Since all these gravity solutions are static black holes, their Euclidean time circle is contractible and we expect a deconfined Polyakov loop, $P_{\beta} \ne 0$.
When IIB gravity provides a good dual description of the SYM we expect the loop (normalized as in~\eqref{eq:Pdefn}) about a cycle in this boundary theory to be non-vanishing if that cycle is contractible when extended into the dual bulk (such as for a cycle about Euclidean time when a horizon exists in the bulk).
Hence, a contractible cycle (i.e., $P_{\beta} \neq 0$) implies a horizon.
Conversely if a cycle is non-contractible in the bulk, we expect the corresponding Wilson loop to vanish.

The distribution of D0~charge on the spatial circle is thought to determine the eigenvalue distribution of $\cP e^{i\oint_L A}$.
In the IIB frame, as the horizon wraps over the spatial circle for the homogeneous black string, this spatial cycle is not contractible in the bulk solution.
Hence at large $N$ we expect spatial confinement, $P_L = 0$, when this homogeneous phase describes the thermodynamics (for $1 \ll \cgrav r_{\beta} < r_L^2$), with the thermal behavior given by eq.~\eqref{eq:D1phase}.
The homogeneity of the horizon is taken to indicate that the eigenvalues of $\cP e^{i\oint_L A}$ are uniformly distributed at large $N$.
On the other hand, upon decreasing the circle size $r_L$ at fixed $r_{\beta}$ we have a first-order transition to the localized phase with thermodynamics given by eq.~\eqref{eq:D0phase}.
Due to the localized horizon, the D0-brane charge is compactly supported on the spatial circle, so we expect the eigenvalue distribution for $\cP e^{i\oint_L A}$ is likewise compactly supported.
This implies spatial deconfinement, $P_L \ne 0$.
The phase transition curve $r_L^2 = \cgrav r_{\beta}$ in the gravity regime, $r_{\beta} \gg 1$, therefore corresponds to a first-order spatial deconfinement transition associated to $P_L$.

We emphasize that we are interested in temperatures and circle sizes where $r_{\beta}$ and $r_L \sim \cO(1)$ in the large-$N$ limit.
If we were to take ultralow temperatures $r_{\beta} \to \infty$ as some sufficiently large positive power of $N$, i.e for $r_{\beta} \sim N$ the theory is thought to enter a conformal phase described by a free orbifold CFT~\cite{Itzhaki:1998dd, Aharony:1999ti}, which we will not explore in this work.

In addition to $r_{\beta} \sim \cO(1)$ we will also consider the high-temperature limit, $r_{\beta} \to 0$ at fixed $r_L$.
Then when $r_{\beta}^3 \ll r_L$ we may integrate out Kaluza--Klein modes on the thermal circle and reduce to a bosonic quantum mechanics (BQM) consisting of the zero modes.
Due to the thermal fermion BCs, this is now the bosonic truncation of the $p = 0$ SYM, as the fermions are projected out in the reduction.
Now the 't~Hooft coupling $\lam_{\text{BQM}}$ is related to the original two-dimensional coupling as $\lam_{\text{BQM}} = \frac{\lam}{\beta}$ and the dynamics implies $\oint A_{\beta} \sim 0$ so that $P_{\beta} \ne 0$.
For small circle size we have $P_L \ne 0$.
This bosonic large-$N$ QM theory has been studied numerically and analytically with the conclusion that as one increases $L$ from zero, the theory confines at $L^3 \lam_{\text{BQM}} \simeq 1.4$~\cite{Kawahara:2007fn, Mandal:2009vz,Azuma:2014cfa}.

\section{Lattice construction}
In this section, we summarize some important features of the four-dimensional lattice theory before proceeding to its dimensional reduction which we use in this work.
The lattice theory based on topologically twisted formulation of the underlying supersymmetric theory can be used to preserve a single \cQ supersymmetry on the lattice as noted in Refs.~\cite{Kaplan:2005ta, Catterall:2007kn, Damgaard:2008pa, Catterall:2014vka}.
These lattice formulations were first obtained from orbifolding and deconstruction methods~\cite{Cohen:2003xe, Cohen:2003qw, Kaplan:2005ta}.
From among different choices of the lattice geometry~\cite{Kaplan:2005ta}, it appears that $A_4^*$ lattice whose \emph{five} basis vectors symmetrically span the four spacetime dimensions is the best choice since it has a high $S_5$ point group symmetry with the dimensions of its low lying irreducible representations matching those of the continuum twisted SO(4) rotation group.
The combination of the \cQ supersymmetry, lattice gauge invariance and the $S_5$ global symmetry suffices to ensure that no new relevant operators are generated by quantum corrections Ref.~\cite{Catterall:2014mha}.

The resultant lattice action can be written down as in Ref.~\cite{Catterall:2014vka, Schaich:2014pda, Catterall:2014vga, Catterall:2015ira, Schaich:2015daa, Schaich:2015ppr, Schaich:2016jus},
\begin{equation}
  \label{eq:lat_act}
  \begin{split}
    S_0 & = \frac{N}{4\lalat} \sum_{\vn} \text{Tr}\left[-\cFb_{ab}(\vn) \cF_{ab}(\vn) + \frac{1}{2}\left(\cDb_a^{(-)}\cU_a(\vn)\right)^2 \right. \\
        & \hspace{5 cm} \left. - \chi_{ab}(\vn) \cD^{(+)}_{[a}\psi_{b]}(\vn) - \eta(\vn) \cDb^{(-)}_a\psi_a(\vn)\right] + S_{\text{cl}}.
  \end{split}
\end{equation}
This leads to the continuum action with $\lam_4 = \lalat / \sqrt{5}$~\cite{Kaplan:2005ta, Catterall:2014vka}.
The presence of an exact lattice supersymmetry allows us to derive an exact expression for the derivative of the partition function with respect to the coupling~\cite{Catterall:2008yz, Catterall:2008dv}, given by
\begin{equation}
  \label{eq:subtraction}
  \vev{s_{\text{Bos}}} = \left(\frac{\vev{S_B^{\text{lat}}}}{V} - \frac{9N^2}{2}\right)
\end{equation}
where $V$ denotes the number of lattice sites and $S_B^{\text{lat}}$ corresponds to the bosonic terms in the lattice action.
This definition of the continuum renormalized $\vev{s_{\text{Bos}}}$ has the property that it vanishes as a consequence of the exact lattice supersymmetry, if periodic (non-thermal) BCs are used.
This quantity will be important when we compare our lattice results to the predictions from the supergravity.

In practice, to stabilize the SU($N$) flat directions of the theory we add to $S_0$ a soft-supersymmetry-breaking scalar potential
\begin{equation}
  \label{eq:single_trace}
  S_{\text{soft}} = \frac{N}{4\lalat} \mu^2 \sum_{\vn,\ a} \Tr{\bigg(\cUb_a(\vn) \cU_a(\vn) - \Ibb_N\bigg)^2}
\end{equation}
with tunable parameter $\mu$.
In the dimensionally reduced system this term is particularly important at low temperatures where the flat directions lead to thermal instabilities~\cite{Catterall:2009xn}.
Exact supersymmetry at $\mu = 0$ ensures that all $\cQ$-breaking counterterms vanish as some power of $\mu$.

The complexification of the gauge field in our construction leads to an enlarged U($N$) = SU($N$) $\otimes$ U(1) gauge invariance.
In the continuum, the U(1) sector decouples from observables in the SU($N$) sector, but this is not automatic at non-zero lattice spacing~\cite{Catterall:2014vka, Schaich:2014pda, Catterall:2014vga}.
To regulate additional flat directions in the U(1) sector, we truncate the theory to remove the U(1) modes from $\cU_a$, making them elements of the group SL($N, \Cbb$) rather than the algebra $\glN$.
In order to maintain SU($N$) gauge invariance it is necessary to keep the fermions in $\glN$, explicitly breaking the lattice supersymmetry that would have related $\cU_a$ to $\psi_a$.
However, by representing the truncated gauge links as $\cU_b = e^{iga \cA_b}$, we can argue that the continuum supersymmetry relating $\cA_a$ and $\psi_a$ is approximately realized in the large-$N$ limit even at non-zero lattice spacing.
This can be understood as follows: fixing the 't~Hooft coupling $\lalat = g^2 N$ as $N \to \infty$, implies $g^2 \to 0$.
Then expanding the exponential produces the desired $\cU_b = \Ibb_N + iga\cA_b$ up to $\cO(ga)$ corrections that vanish as $N \to \infty$ even at non-zero lattice spacing $a$.
Therefore, one can sacrifice the exact supersymmetry in lieu of controlling the instabilities with just a small mass term (which turns out to be crucial for understanding the dual thermodynamics) by simulating SU(N) theory rather than full U(N).
In addition, for the dimensionally reduced lattice theory to correctly reproduce the physics of the continuum theory requires $\Tr{\varphi_i} \approx N$ in the reduced directions $i = y$ and $z$.
This corresponds to broken center symmetries in those two directions.
We ensure this by adding another soft-$\cQ$-breaking term to the lattice action,
\begin{equation}
  \label{eq:center}
  S_{\text{center}} = -\frac{N}{4\lalat} c_W^2 \sum_{\vn,\ i = y, z} 2\text{ReTr}\bigg[\varphi_i(\vn) + \varphi_i^{-1}(\vn)\bigg].
\end{equation}
This is gauge invariant since $\varphi_i(\vn)$ transform as site fields.
In this work we use either $c_W^2 = \mu^2$ or $c_W^2 = 0$, again extrapolating $\mu^2 \to 0$ in the former case \footnote{The code used for this work is available to download {\tt\href{https://github.com/daschaich/susy/releases/tag/1709.07025}{here}}}.

\section{Numerical results \& Conclusions}
\begin{figure}[tbp]
  \includegraphics[height=4.5cm]{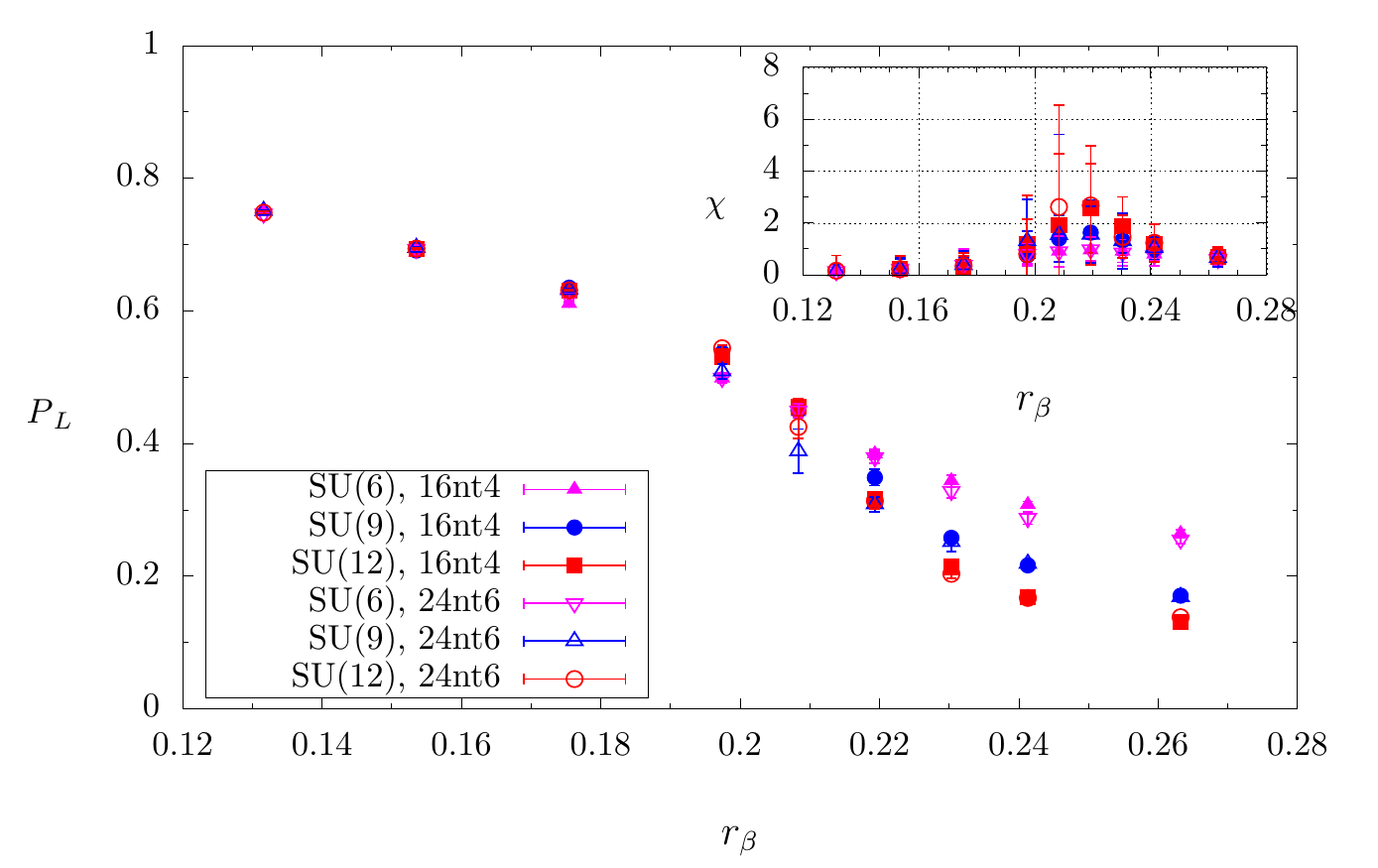}\hfill \includegraphics[height=4.5cm]{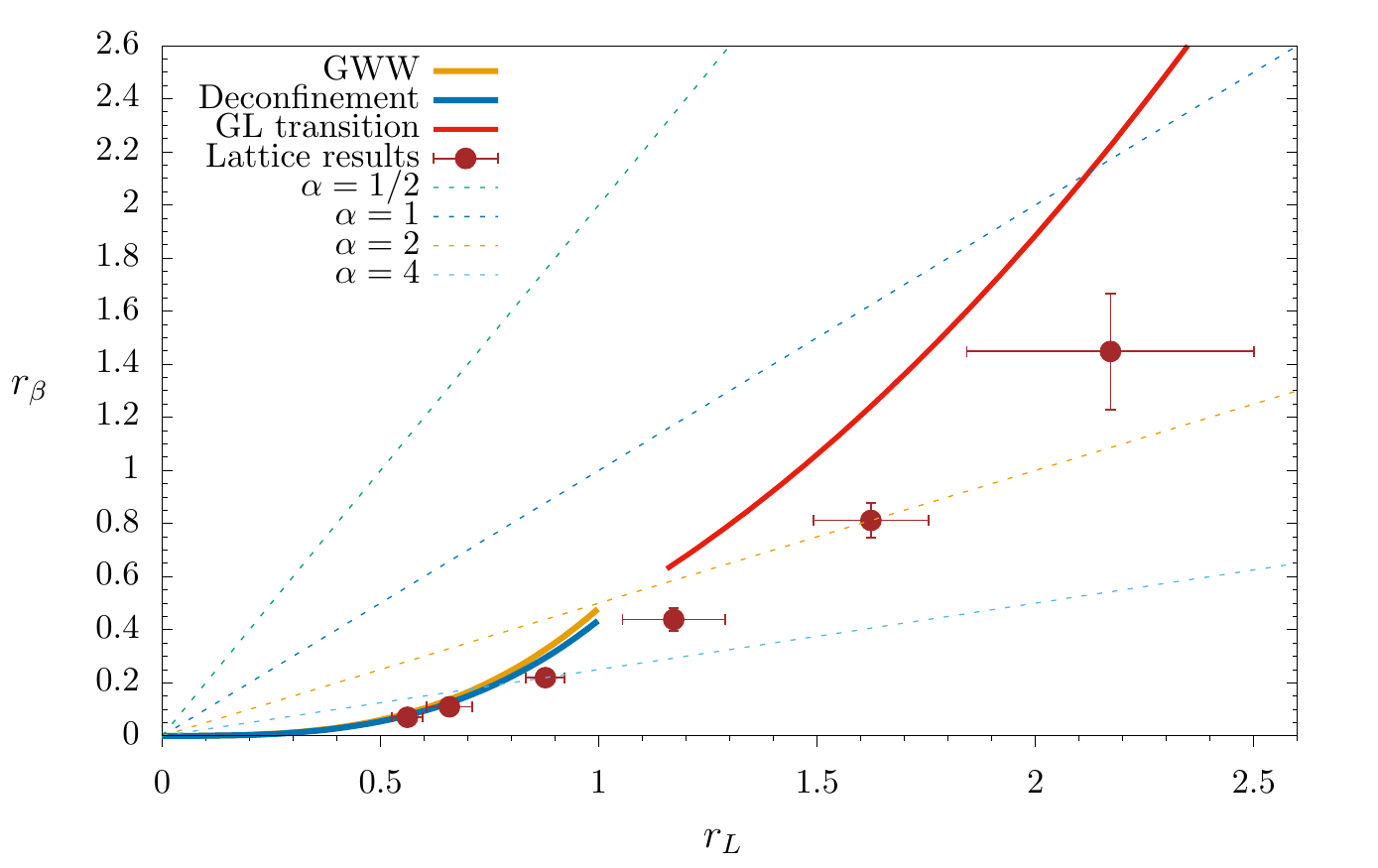}
  \caption{\label{fig:crit}Left : The average magnitude of the Wilson line $P_L$ vs.~$r_{\beta}$ for $\al = 4$. The inset shows the corresponding susceptibility,  $\chi = \vev{\left| \Tr{\cP e^{i\oint_L A}} \right|^2} - \vev{\left| \Tr{\cP e^{i\oint_L A}} \right|}^2 $. Right : Similar analysis for series of aspect ratios, $\alpha$ showing the corresponding locations of the deconfinement transition. }
\end{figure}

\begin{figure}[htbp]
  \includegraphics[height=4.5cm]{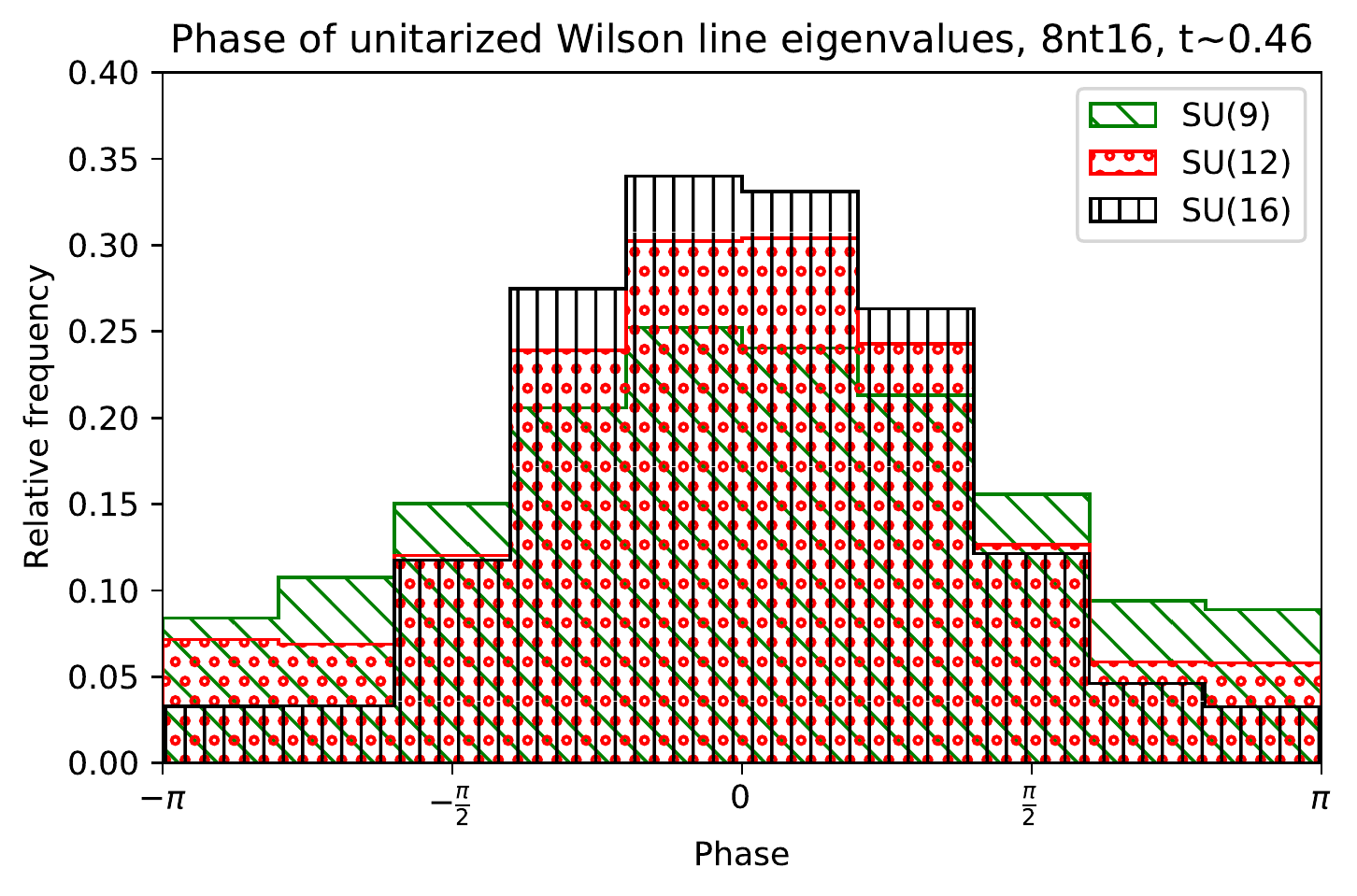}\hfill \includegraphics[height=4.5cm]{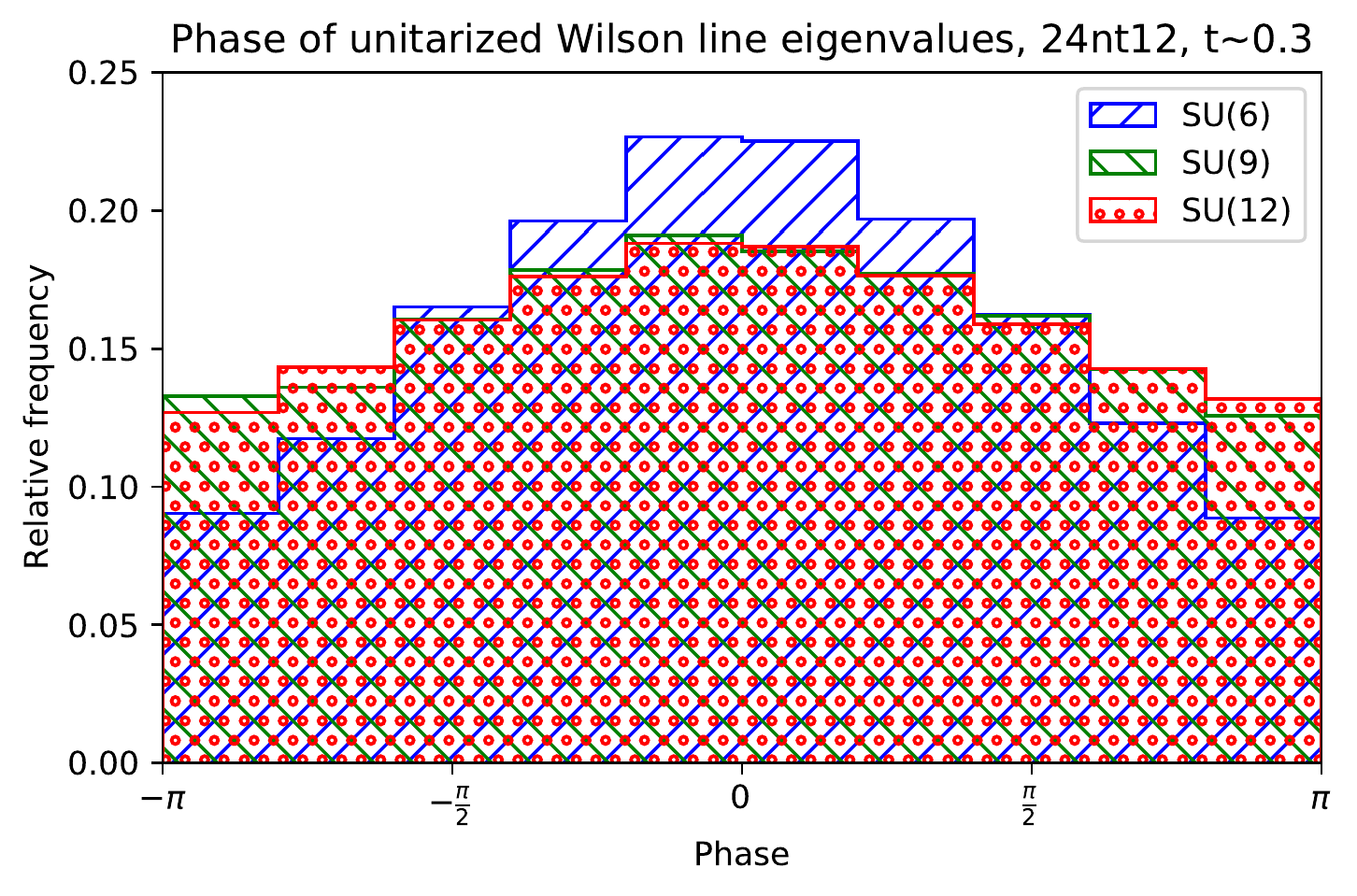}
  \caption{\label{fig:WLeig} Left : Distributions of Wilson line eigenvalue phases for $8\times 16$ lattices at $t \approx 0.46$ with $\mu^2 \approx 0.004$.  The intermediate distributions, which become more compact as $N$ increases, are consistent with expectations from the D0~phase of the gravity dual. Right : Distributions of Wilson line eigenvalue phases for $24\times 12$ lattices at $t \approx 0.33$ with $\mu^2 \approx 0.007$.  The extended distributions, which become more uniform as $N$ increases, correspond to the D1~phase of the gravity dual. }
\end{figure}

\begin{figure}[htbp]
  \includegraphics[height=4.5cm]{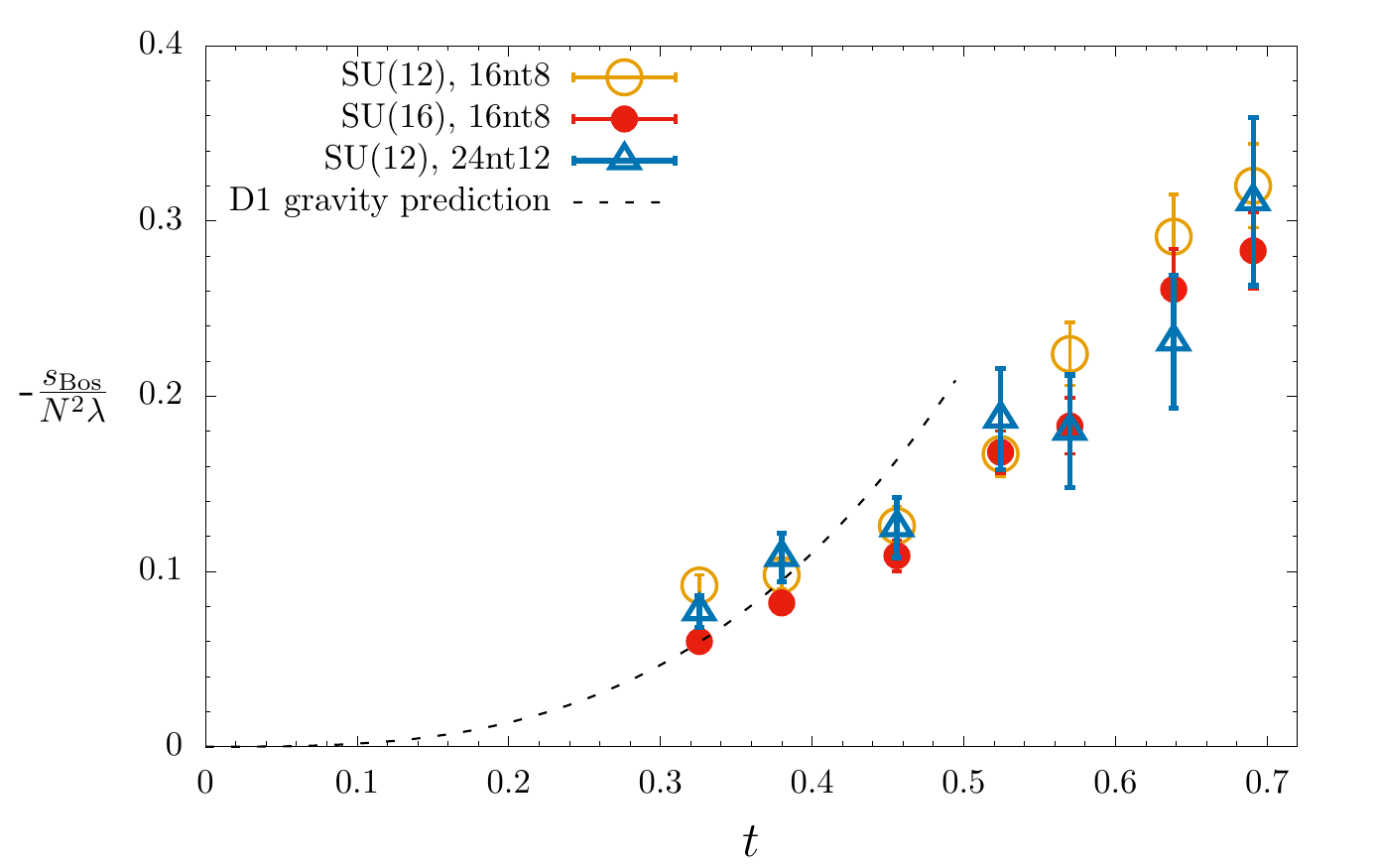}\hfill \includegraphics[height=4.5cm]{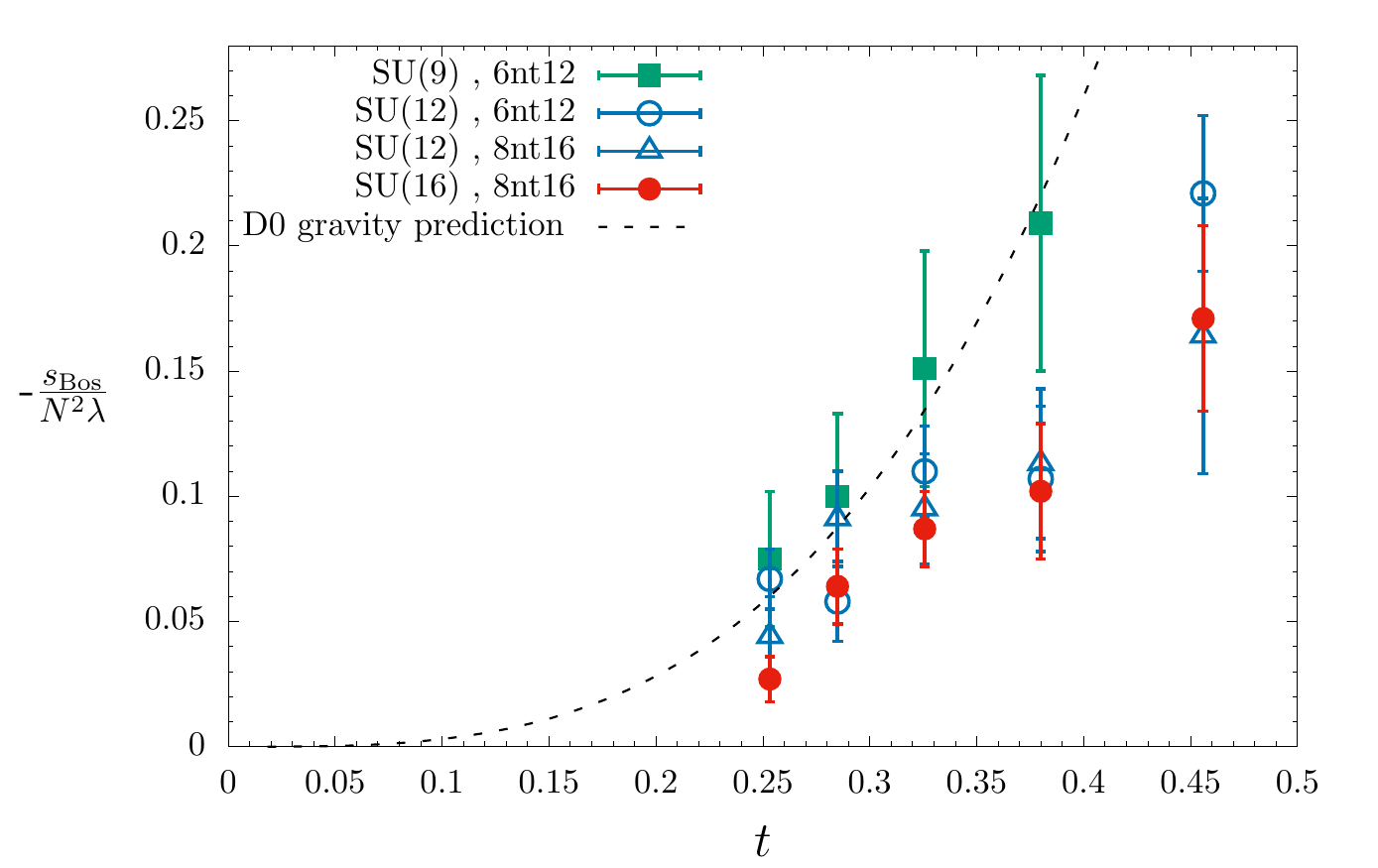}
  \caption{\label{fig:alpha2}Left : Bosonic action density vs.\ temperature $t$ for $\al = 2$ with $N = 12$ and 16. Right : Bosonic action density vs.\ temperature $t$ for $\al = 1 / 2$ with $N = 9$, 12 and 16.}
\end{figure}

In figure~\ref{fig:crit} (left), we show the jackknife average magnitude of the Wilson line $P_L$ vs.~$r_{\beta}$ for $\al = 4$, along with the corresponding susceptibility.
The results indicate a large-$N$ transition at $t_c = 4.6(2)$ separating a spatially deconfined phase with $P_L \ne 0$ at small $r_{\beta}$ (high temperatures) from a spatially confined phase at large $r_{\beta}$ (low temperatures) where $P_L \to 0$ as $N \to \infty$.

We see a transition between localized and homogeneous phases, as shown in the right panel of Fig.~\ref{fig:crit}.
The parametric form ($r_L^2 = \cgrav r_{\beta}$) agrees with the expectations from the gravity calculations.
However, we have found that determination of this phase transition for $\al \le 1$ is not possible even at $N=12$.
Moving on to the detailed thermodynamic behavior, in the left panel of Fig.~\ref{fig:alpha2} we show the action density versus $t$ for $\al = 2$ lattice volumes $16\times 8$ and $24\times 12$ with gauge groups SU(12) and SU(16). For sufficiently large~$N$ the low-temperature regime of this plot should correspond to the D1~phase of the gravity dual.
This is confirmed by our lattice results, which lie close to the D1-gravity prediction at low temperatures. In the right panel of Fig.~\ref{fig:alpha2} we show the action density versus $t$ for $\al = 1/2$ lattice volumes $6\times 12$ and $8\times 16$ with gauge groups SU(9), SU(12), and SU(16) with dashed curves showing the predictions for the localized D0~phase from gravity. 

In summary, our numerical results for the phase diagram of the two-dimensional SYM are consistent with the expectations from the dual gravity theory.
We see a spatially confined phase where the eigenvalues of the Wilson line are uniformly distributed around the unit circle, as expected for a homogeneous black-string phase.
This is separated by a first-order transition from a confined phase having localized eigenvalue distribution corresponding to a D0~phase. 


\vspace{12 pt}
\noindent {\sc Acknowledgments:}~We thank Krai Cheamsawat, Joel Giedt and Anosh Joseph for helpful conversations.
This work was supported by the U.S.~Department of Energy (DOE), Office of Science, Office of High Energy Physics, under Award Number DE-SC0009998 (SC, RGJ). 
Numerical calculations were carried out on the DOE-funded USQCD facilities at Fermilab, and at the San Diego Computing Center through XSEDE supported by National Science Foundation Grant No.~ACI-1053575.

\raggedright
\bibliography{Lattice2017_50_JHA}
\end{document}